# Impulsive rotational Raman scattering of $N_2$ by a remote "air laser" in femtosecond laser filament

Jielei Ni,[1] Wei Chu,[1] Haisu Zhang,[1,3] Bin Zeng,[1] Jinping Yao,[1] Guihua Li,[1,3] Chenrui Jing,[1,3] Hongqiang Xie,[1,3] Huailiang Xu,[2,†] Ya Cheng,[1,‡] and Zhizhan Xu[1,*]

[1] State Key Laboratory of High Field Laser Physics, Shanghai Institute of Optics and Fine Mechanics, Chinese Academy of Sciences, P.O. Box 800-211, Shanghai 201800
[2] State Key Laboratory on Integrated Optoelectronics, College of Electronic Science and Engineering, Jilin University, Changchun 130012, China
[3] Graduate School of the Chinese Academics of Sciences, Beijing 100039, China

† huailiang@jlu.edu.cn

‡ ya.cheng@siom.ac.cn

* zzxu@mail.shcnc.ac.cna

We report on experimental realization of impulsive rotational Raman scattering from neutral nitrogen molecules in a femtosecond laser filament using an intense self-induced white-light seeding "air laser" generated during the filamentation of an 800 nm Ti: Sapphire laser in nitrogen gas. The impulsive rotational Raman fingerprint signals are observed with a maximum conversion efficiency of ~0.8%. Our observation provides a promising way of remote identification and location of chemical species in atmosphere by rotational Raman scattering of molecules.

Because of the unique ability of kilometer range nonlinear propagation [1,2] without perturbation in adverse conditions [3,4], femtosecond laser filament has been regarded as one of the most attractive potential tools for remote applications in atmosphere [5,6]. Inside the filament core, optical intensity is clamped at ~$5\times10^{13}$ W/cm$^2$ due to the dynamic balance between Kerr self-focusing and defocusing of plasma induced by multiphoton/tunnel ionization of air molecules [7,8]. With such high intensity, ionization and fragmentation of molecules would occur, resulting in characteristic fingerprint emissions that can be used for sensing pollutants in air [6]. However, such filament-based spectroscopic techniques suffer from limited signal-to-noise ratio due to the self-induced white light during the filamentation, which spans from ultraviolet to the infrared, especially when the filament is long [9]. Nevertheless, this drawback may be overcome by using a so-called "air lasing" phenomena in filaments to enhance the intensity of the characteristic emissions via amplified spontaneous emission (ASE) [10-12] or stimulated seed amplification [13-15]. Currently, such air lasing phenomena have been demonstrated for the gas targets including nitrogen gas [10,13,15], water vapor [11], air-hydrocarbons gas mixture (i.e. $CH_4$, $C_2H_2$, and $C_2H_4$)[12], and carbon dioxide gas [14]. In particular, it was recently demonstrated that self-induced white-light seeding air lasers can be generated remotely by using only one 800 nm Ti: Sapphire beam [16-18], and the self-lasing pulse maintains the optical properties of the 800 nm pump laser pulse, such as coherence and polarization. Interestingly, other than providing itself as a characteristic signal of high signal-to-noise ratio, the air laser could also be an ideal intense, narrow-bandwidth laser source as we will show in this paper, for identifying air components through optical spectroscopy such as Raman scattering.

Previously, impulsive Raman scattering has been observed for a variety of gas-phase targets [19-21] using femtosecond pump-probe schemes, in which a femtosecond laser pulse is introduced for impulsive vibrational/rotational excitation of molecules, and a sub-picosecond laser pulse is used to produce Raman spectra subsequently. However, such method encounters the difficulty to achieve perfect spatial and temporal overlaps between the two beams over a long distance. In this Letter, we show for the first time that, impulsive rotational Raman scattering of molecules can be remotely initiated inside a femtosecond filament by using the self-induced "air laser" which can provide a narrow-bandwidth, intense sub-10 ps laser pulse with the interaction distance as far as the filament [16]. This scheme provides two unique advantages. First, the pump and the probe pulses naturally overlap in both time and space, since the air laser probe pulse is generated in the filament induced by the pump laser. Second, the femtosecond laser filamentation process offers the potential to realize the Raman fingerprinting in a remote distance. Both these advantages are highly desirable for remote atmospheric sensing.

The experiment was performed using ~40 fs (FWHM), 800 nm laser pulses with a pulse energy of 15 mJ from a commercial 800 nm Ti: sapphire laser system (Legend Elite Cryo PA, Coherent, Inc.) at a repetition rate of 1 kHz. The beam diameter was measured to be ~8.8 mm (1/e$^2$). As shown in Fig. 1, the laser beam was focused into a gas chamber filled with nitrogen gas at a pressure of 1 atm by a fused silica lens (L1) with a focal length of 40 cm, forming a filament in the gas chamber. The forward light signals containing the supercontinuum, the lasing emission at wavelength ~428 nm, and the rotational Raman scattering were firstly attenuated and reflected by a glass plate, and then collimated by a fused silica lens with a focal length of 40 cm. The lasing emission at wavelength ~428 nm and the Raman signals were isolated from the supercontinuum by a transmission dielectric filter with the central wavelength at 428 nm and the bandpass of 10 nm. Then they were further attenuated by 90% with a neutral density filter. Finally, the light signals were focused onto the entrance slit of an imaging grating spectrometer (Shamrock 303i, Andor) by a fused silica lens with a focal length of 13 cm. In the measurement, the width of the entrance slit of the spectrometer was fixed at 100 μm. The fluorescence signals were also collected from the side of the filament by a lens with a focal length of 3.6 cm and detected by the same spectrometer (Shamrock 303i, Andor).

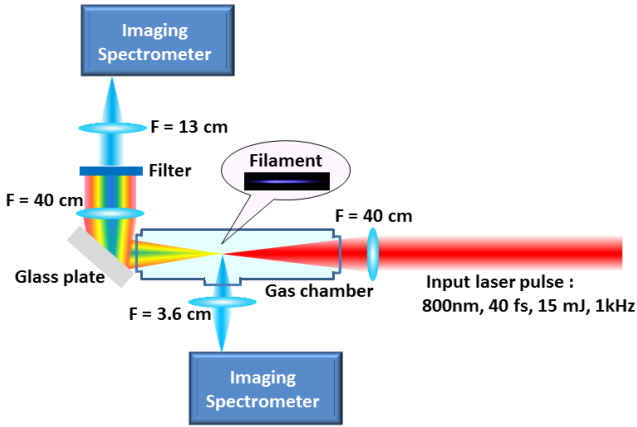

Fig. 1. (Color online) Schematic of the experimental setup.

Figure 2(a) illustrates the forward spectrum recorded with 1 second exposure time, corresponding to 1000 laser shots. It can be clearly seen in Fig. 2(a) that there exists a strong, narrow-bandwidth lasing emission centered at 427.8 nm, which is attributed to the self-induced white-light seeded lasing emission between the vibrational transition of the $B^2\Sigma_u^+$ ($v = 0$) and $X^2\Sigma_g^+$ ($v = 1$) states of $N_2^+$ [13]. Besides the strong laser line at 427.8 nm, six weak and closely spaced peaks can also be observed in the forward spectrum of Fig. 2(a) on the blue side of the 427.8 nm laser line and five on the red side. For comparison, we show the spectrum in Fig. 2(b) recorded from the side of the filament with a 20 second exposure time. It can be seen that the spectrum shown in Fig. 2(b) features typical fluorescence emissions with the band head at 428.7 nm. In addition, similar to those shown in Fig. 2(a), weak, closely spaced peaks can be observed in Fig. 2(b) on the blue side of the fluorescence signal at 427.8 nm. Note that the peak located at 423.6 nm in Fig. 2(b) is the fluorescence resulted from the vibrational transition of the $B^2\Sigma_u^+$ ($v = 1$) and $X^2\Sigma_g^+$ ($v = 2$) states. However, different from the spectrum of Fig. 2(a), those peaks shown on the red side of the laser line in Fig. 2(a) are missing (i.e., unobservable in our experiment) in Fig. 2(b).

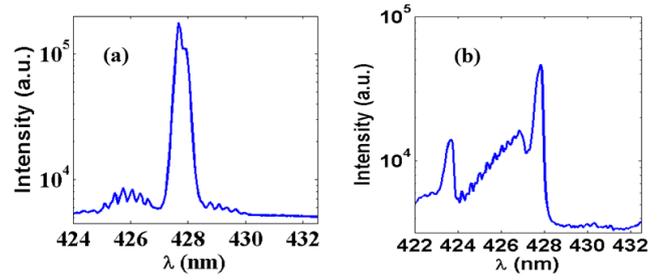

Fig. 2. (Color online) (a) Measured lasing spectrum in the forward direction; (b) Measured fluorescence spectrum from the side of the filament.

To reveal the mechanisms behind these peaks, we firstly converted the spectrum of Fig. 2(a) into the wavenumber scale with the laser line of 427.8 nm as the origin, as shown in Fig. 3(a), in which the shift of different frequency peaks in Fig. 2(a) with respect to the 427.8 nm laser line can be obtained based on the relation of $|1/\lambda_{427.8} - 1/\lambda|$, where $\lambda_{427.8}$ denotes the wavelength of the lasing emission peak, and $\lambda$ stands for the wavelength of the other signals. It can be seen in Fig. 3(a) that the peaks located in the spectrum with the peak value larger than 50 cm$^{-1}$ can be well distinguished and spectrally separated. Furthermore, it can be observed in Fig. 3(a) that the frequency difference between the two successive peaks is nearly constant with the value of ~16 cm$^{-1}$. According to the Raman shift, $\Delta v = B_e (4J + 6)$, with the rotational constant of $B_e = 1.998$ cm$^{-1}$ for neutral nitrogen molecules in the ground state and the rotational quantum number J, the S-branch (J→J+2) rotational Raman lines of S(6), S(8), S(10), S(12) and S(14) can be calculated to be 59.94, 75.92, 91.91, 107.89, and 123.88 cm$^{-1}$, respectively, as shown by the red circles in Fig. 3(a), which are all in good agreement with the measured shifts of 59.8 cm$^{-1}$, 76.6 cm$^{-1}$, 91.0 cm$^{-1}$, 107.5 cm$^{-1}$, 124.4 cm$^{-1}$, respectively. In addition, the S-branch rotational Raman lines of S(0), S(2) and S(4) can be calculated, which are located at 11.99, 27.97, and 43.96 cm$^{-1}$, respectively. Although these three measured Raman peaks in Fig. 3(a) are masked by the strong 427.8 nm laser signal, it can be observed that a good agreement between the experimental and theoretical positions for the S(0) and S(4) peaks can be realized, as shown by the dashed lines in Fig. 3(a). As a result, we attributed these side peaks to impulsive rotational Raman scattering of neutral nitrogen molecules triggered by the self-induced white light seeding "air laser" at ~428 nm in the filament, whose mechanism is similar to that described in [19]. This observation also indicates that, despite the high input energy used in this experiment, the overwhelming majority of gas inside the filament may still be the neutral nitrogen molecules due to the intensity clamping. From the measured Raman peaks shown in Figs. 2(a) and 3(a), a conversion efficiency of ~ 0.8% can be evaluated. In addition, we characterize the polarization property of the Raman peaks on the red side of the 427.8 nm laser by placing a Glan-Taylor polarizer prism in front of the spectrometer. It is found that these peaks are nearly linearly polarized with the direction parallel to that of the laser line, as demonstrated in Fig. 3(b).

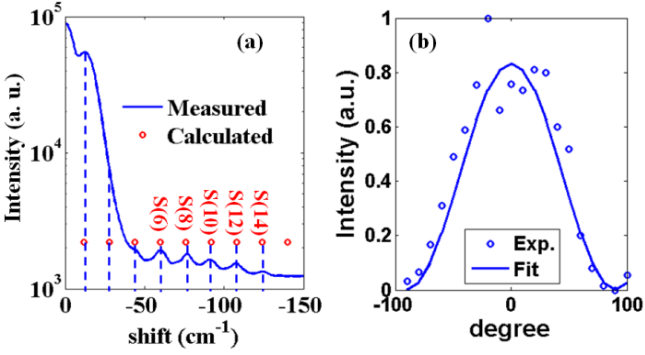

Fig. 3. (Color online) (a) Measured Raman spectrum (blue solid) induced by intense lasing emission at 428 nm, and calculated Raman peaks in neutral nitrogen with a rotational constant $B_e$ = 1.998 cm$^{-1}$ (red circle). (b) Measured polarization property of the Raman signals (blue circle) fitted by $\cos^2\theta$ (blue solid).

Next, we examine the peaks on the blue side of the 427.8 nm laser line shown in Fig. 2(a). However, different from the equidistant peaks of the Raman scattering mentioned above, the spacing between the successive two peaks on the blue side of the 427.8 nm laser line become larger toward shorter wavelength. Such behavior is similar to that of the R-branch transition of $N_2^+$ described in Refs. [22,23]. Taking the anharmonicity and the interaction of rotation and vibration into account, the wave number of the R-branch transition between $B^2\Sigma_u^+$ ($v$ = 0) and $X^2\Sigma_g^+$ ($v$ = 1) states can be determined by [24],

$R(J) = v_{00} - (\omega_e - 2\omega_e\chi_e) + 2B_v' + (3B_v' - B_v'')J + (B_v' - B_v'')J^2$,

where $B_v' = B_e' - \alpha_e'/2$ with $B_e'$ = 2.083 cm$^{-1}$ and $\alpha_e'$ = 0.0195 cm$^{-1}$ is the rotational constant of the upper state $B^2\Sigma_u^+$ ($v$ = 0), $B_v'' = B_e'' - 3\alpha_e''/2$ with $B_e''$ = 1.932 cm$^{-1}$ and $\alpha_e''$ = 0.020 cm$^{-1}$ is the rotational constant of the lower state $X^2\Sigma_g^+$ ($v$ = 1), and $v_{00}$ = 25566 cm$^{-1}$, $\omega_e$ = 2207.19 cm$^{-1}$ and $\omega_e\chi_e$ = 16.136 cm$^{-1}$ are given in Ref. [24]. As a result, the calculated R-branch transitions between the $B^2\Sigma_u^+$ ($v$ = 0) and $X^2\Sigma_g^+$ ($v$ = 1) states of $N_2^+$ are shown by the black circles in Fig. 4(a). These results agree well with the measured R-branch transitions peaks of the fluorescence spectrum, which is the higher resolution spectrum in fig. 2(b).

Therefore, we calculate both the O-branch (J→J-2) rotational Raman peaks and R-branch transitions as described above, and compare them with the measured blue-side peaks of the 427.8 nm laser line in the forward spectrum, as shown in Fig. 4(b). It can be seen that although the calculated R-branch transitions locate closely to the positions of O-branch (J→J-2) rotational Raman transitions, noticeable difference can be observed between the two groups of peaks. We gather that the difference is caused by the contamination by the rotational Raman scattering, namely, the peaks in the blue side are contributed by both the R-branch transition and the rotational Raman scattering.

To summarize, we have observed impulsive rotational Raman scattering of neutral nitrogen gas in a femtosecond laser filament with a conversion efficiency of ~0.8%. Typically, femtosecond-laser-pulse-induced Raman scattering shows a broad band, whereas the Raman scattering spectrum observed in our experiment are spectrally resolved with a high resolution with the air laser serving as the probe pulses. Our observation indicates that such remote air laser generated in femtosecond laser filaments could provide a promising narrow-bandwidth source for remote atmospheric spectroscopy applications.

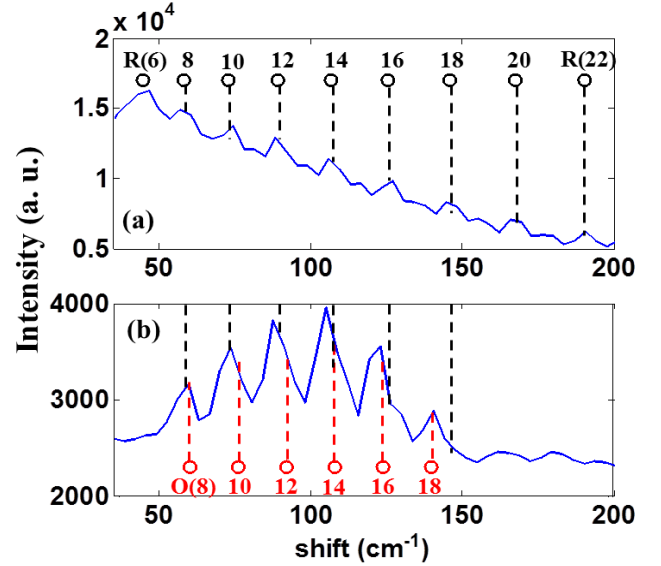

Fig. 4. (Color online) (a) Measured fluorescence spectrum of the R-branch of $B^2\Sigma_u^+$ ($v$ = 0) - $X^2\Sigma_g^+$ ($v$ = 1) states from the side of the filament, in which R-branch is labelled with different J numbers in the lower states. (b) The blue side spectrum of the lasing emission of $B^2\Sigma_u^+$ ($v$ = 0) - $X^2\Sigma_g^+$ ($v$ = 1) states measured in the forward direction.

This work is supported in part by the National Basic Research Program of China (Grants No. 2011CB808100 and No. 2014CB921300), National Natural Science Foundation of China (Grants No. 11127901, No. 11134010, No. 11204332, No. 11304330, and No.11004209, No. 61235003), the Program of Shanghai Subject Chief Scientist (11XD1405500), the Open Fund of the State Key Laboratory of High Field Laser Physics (SIOM), and the Fundamental Research Funds of Jilin University.